\documentclass{ws-proc9x6}

\usepackage{epsfig}
\epsfclipon
\def\beqa{\begin{eqnarray}}
\def\eeqa{\end{eqnarray}}
\def\beq{\begin{equation}}
\def\eeq{\end{equation}}
\def\subp{+}
\def\subm{-}
\def\subo{{(1)}}
\def\subt{{(2)}}

\def\lsim{\mbox{\raisebox{-.6ex}{~$\stackrel{<}{\sim}$~}}}

\def\alphp{\alpha'}

\begin{document}


\title{Tachyon Defect Formation and Reheating in Brane-Antibrane Inflation}

\author{Neil Barnaby \footnote{ \it presenter}}
\address{McGill University, 3600 University St.\
Montr\'eal, Qu\'ebec H3A 2T8, Canada \\
E-mail: barnaby@physics.mcgill.ca}  

\author{James M. Cline}
\address{McGill University, 3600 University St.\
Montr\'eal, Qu\'ebec H3A 2T8, Canada \\
E-mail: jcline@physics.mcgill.ca}  


\maketitle

\abstracts{
We study analytically the dynamical formation of lower dimensional branes at the endpoint of brane-antibrane
inflation through the condensation of topological defects of the tachyon field which describes the 
instability of the initial state.  We then use this information to quantify the efficiency of the reheating
which is due to the coupling of time dependent tachyon background to massless gauge fields which will be 
localized on the final state branes.  We improve upon previous estimates indicating that this can be an
efficient reheating mechanism for observers on the brane.
}


\section{Brane-Antibrane Inflation and the Model of Reheating}

We begin by briefly describing brane-antibrane inflation and the scenario we have in mind for reheating,
see \cite{Reheating2} for a more detailed discussion and references.
In the simplest model of D-brane inflation a parallel brane and antibrane begin with some separation between 
them in one of the extra dimensions.  There is an attractive force between the branes, primarily due to the
exchange of massless bulk modes.  The branes move towards each other with the inter-brane separation playing
the role of the inflaton and the non-vanishing potential energy between the branes driving inflation.  Once 
the branes have reached a critical separation one of the stretched string modes between the branes, $T$, 
becomes tachyonic which provides a natural mechanism to end inflation
since the brane-antibrane pair then become unstable to annihilate.  The annihilation of the initial state
brane-antibrane pair is described by the rolling of the tachyon field from the unstable maximum of its 
potential to the degenerate vacua.  The tachyon field may form topological defects throught the Kibble 
mechanism so that $T=0$ stays fixed at the core of the defect.  These defects are known to be consistent 
descriptions of lower dimensional branes and hence the condensation of tachyon defects describes the 
formation of lower dimensional branes at the endpoint of brane-antibrane inflation.

Our interest in studying the endpoint of D-brane inflation is motivated by questions about the efficiency
of the reheating.  It is possible that the energy density liberated from the brane collisions will be 
converted into closed string states (ultimately gravitons) and not necessarily into visible radiation.
The formation of tachyon defects at then endpoint of D-brane inflation is a dynamical process where the 
tachyon couples to gauge fields which will be localised on the descendant brane.  It is thus expected
that some radiation will be produced by the rolling of the tachyon and the problem of reheating becomes
quantitative: can this effect be efficient enough to strongly deplete the energy density of the tachyon 
fluid so that the universe starts out dominated by radiation rather than cold dark matter?  This question
has been addressed in a similar context in \cite{Reheating1}.  

To study the reheating quantitatively we consider a model in which our universe is a D$3$-brane which 
is the decay product of the annihilation of a D$5$-D$\bar{5}$ pair.  We will begin by studying tachyon 
defect formation in an effective $6$D theory with the two extra dimensions compactified on a sphere.  In 
this picture our universe is a tachyon vortex located at the north pole of the $2$-sphere.  Conservation of 
RR charge requires that a D$\bar{3}$-brane also be created in the annihilation which we place at the south 
pole of the $2$-sphere, to preserve azimuthal symmetry.  Throughout the analysis we will study the formation
of only one defect since both defects are identical and the solutions may be matched along the hemisphere.  
Furthermore, since the defects will be highly localized at the poles we will ignore the compact topology of 
the extra dimensions in solving for the vortex background.

\section{Effective Field Theory on the Brane-Antibrane Pair \label{vortex}}

We will work with an effective field theory proposed by Sen \cite{Sen} for the tachyon on the brane-antibrane
pair.  The field content for this system is a complex tachyon field $T$, massless gauge fields 
$A_M^{\subo}$, $A_M^\subt$
\footnote{Upper case roman indices $\{M,N\}$ 
run over the full space-time coordinates $\{0,1,\cdots,p\}$.  We also define
greek indices $\{\mu,\nu\}$ which run over the coordinates $\{0,1,2\}$ on which the vortex solutions depend 
and hatted greek indices $\{\hat{\mu},\hat{\nu}\}$ which  run over the 
spatial coordinates parallel to the vortex $\{2,3,\cdots,p\}$. Here $p=6$ for a vortex which
describes a 3-brane.}
and scalar fields $Y^I_\subo$, $Y^I_\subt$ corresponding to the 
transverse fluctuations of the branes.  The index $(i)=(1),(2)$ labels
which of the original branes (actually the brane or the antibrane) the field is associated with.
The effective action is:
\begin{equation}
\label{action}
  S = -\int V(T,Y^I_\subo-Y^I_\subt) \left( 
  \sqrt{-\det M^\subo}  \right. 
  + \left. \sqrt{-\det M^\subt} \,  \right)
  \, d^{\,p+1}\! x 
\end{equation}
where 
\[
  M^{(i)}_{MN} = g_{MN} + \alphp F^{(i)}_{MN} + \partial_M Y^I_{(i)}\partial_N Y^I_{(i)}
  + \frac{1}{2}D_M T D_N T^\ast +  \frac{1}{2}D_M T^\ast D_N T,
\]
\[
  F^{(i)}_{MN} = \partial_M A^{(i)}_N - \partial_N A^{(i)}_M, \hspace{5mm}
  D_M = \partial_M -iA^\subo_M+iA^\subt_M.
\]
For the remainder of this paper we will ignore the transverse
scalars and choose $V(T,0)=V(T)=\tau_p \exp \left(-{|T|^2}/{a^2} \right)$ where 
$a$ is chosen so that the static singular vortex solutions of the theory 
(\ref{action}) have the correct tension to be interpreted as codimension 2 D-branes according the 
the normalization proposed in \cite{Sen}.

As a first step we would like to study the dynamics of formation of a tachyon vortex 
without considering the coupling to the photon.
To construct vortex solutions we ignore the transverse scalars and take the remaining fields to 
depend only on the polar coordinates $x^\mu=(x^0,x^1,x^2)=(t,r,\theta)$ with metric 
$g_{\mu\nu} dx^\mu dx^\nu = -dt^2 + dr^2 + r^2 d\theta^2$
\footnote{Recall that we are ignoring the compact topology of the extra dimensions.  The radial coordinate
$r$ measures distance along the surface of the $2$-sphere from the north/south pole for the 
vortex/antivortex solution.}.  Since the vortex solution 
should have azimuthal symmetry we make the ansatz:
\begin{equation}
\label{ansatz}
  T(t,r,\theta)=e^{i\theta}f(t,r), \hspace{5mm} 
                A_\theta^\subo=-A_\theta^\subt = \frac{1}{2} g(t,r)
\end{equation}
with all other components of $A^{(i)}_M$ vanishing. 

\subsection{Solutions Near the Core of the Defect}

To analytically study the dynamics near the core of the vortex, $r=0$, we make the ansatz:
\begin{equation}
\label{smallfansatz}
  f(t,r) \cong p(t)r, \hspace{5mm}  g(t,r) \cong q(t)r
\end{equation}
for small $r$.  Plugging (\ref{smallfansatz}) into the equations of motion which follow from (\ref{action})
and dropping terms which are subleading in $r$ yields a set of coupled ODEs for $p(t)$ and $q(t)$.  One finds
that both $p(t)$ and $q(t)$ and singular in finite time  $t_c$, and that near $t_c$ we can approximate
$p(t)$ and $q(t)$ by
\begin{equation}
\label{divergences}
  p(t) = \frac{p_0}{t_c-t}, \hspace{5mm} q(t)= \frac{q_0}{t_c-t}.
\end{equation}

\subsection{Vacuum Solutions}

Away from the core of the defect the fields $f(t,r)$ and $g(t,r)$ tend towards their vacuum values 
$f \rightarrow \infty$ and $g \rightarrow 1$.  To study the dynamics in this regime we make the ansatz
\[
  g(t,r)=1-\varepsilon \sigma(t,r)
\]
and work only to leading order in $\varepsilon$.  One finds that the tachyon field must obey the eikonal
equation: $-(\partial_t f)^2+(\partial_r f)^2+1=0$.
To circumvent the problem caustic formation in the bulk \cite{Caustics} we take the simple solution 
$f(t,r) = t$.  Given this solution any function $\sigma(t,r) = \sigma(r)$ satisfying the necessary
boundary condition $\sigma(r \rightarrow \infty) \rightarrow 0$ will generate a solution.

\subsection{Stress-Energy Tensor}

The finite time divergence in the slope of the tachyon field (\ref{divergences}) will lead to the finite time
formation of a singularity in the stress energy tensor of the vortex.  This divergence has the form of a 
delta function in which $t_c-t$ plays the role of the small parameter which regularizes the delta function.
In the limit as $t \rightarrow t_c$ on finds that the energy momentum tensor for the vortex is identical
to that of a D$(p-2)$-brane
\begin{eqnarray}
  T^{00} &=& \tau_{p-2}\,\delta(r\cos\theta)\,\delta(r\sin\theta) + 2r\Sigma(r)  \nonumber \\
  T^{11} &=& T^{22} = 0 \label{finalTmunu} \\
  T^{\hat{\mu}\hat{\nu}} &=&  -\delta^{\hat{\mu}\hat{\nu}} 
                               \tau_{p-2}\delta(r\cos\theta)\,\delta(r\sin\theta) \nonumber
\end{eqnarray}
where $\Sigma(r)$ is subject to mild constraints from conservation of energy but is otherwise arbitrary.
The extra bulk energy density $2 r \Sigma(r)$ corresponds to tachyon matter rolling  in the bulk.

\section{Inclusion of Massless Gauge Fields}

We consider now the coupling of massless gauge fields to the time dependent tachyon vortex background
described above.  There are two gauge fields in the
problem: $A^{M}_\subo$ and $A^{M}_\subt$, or equivalently $A^{M}_\subp=A^{M}_\subo+A^{M}_\subt$ and
$A^{M}_\subm=A^{M}_\subo-A^{M}_\subt$, which have different couplings to the tachyon.  We have already
shown that $A^{\mu}_\subm$ is the field which condenses in the vortex, hence its associated gauge symmetry is
spontaneously broken. For reheating it is thus $A^{\hat{\mu}}_\subp$ which most closely resembles the
Standard Model photon.  We will ignore fluctuations of the  heavy fields $A^{\hat{\mu}}_\subm,$
$A^{\mu}_\subp$, and $A^{\mu}_\subm$, keeping only the background solution for $A^{\mu}_\subm$ (which was
given in section \ref{vortex}), and the fluctuations of the photon  $A^{\hat{\mu}}_\subp$.
To compute the production of 
photons in the time-dependent background, we want to expand the action
(\ref{action}) to quadratic order in $A^{\hat{\mu}}_\subp$.  The result is
\begin{equation}
\label{effectiveaction}
   S=-\frac{\alphp^2}{4} \int V(f)  
    \sqrt{-G} G^{MN}\delta^{\hat{\mu}\hat{\nu}}\partial_M A_{\hat{\mu}}^\subp \partial_N A_{\hat{\nu}}^\subp
     \, d^{p+1} x
\end{equation}
where the ``tachyon effective metric'' $G_{MN}$ can be related to the vortex stress-energy tensor
as $r  T^{MN} = - 2 V(f) \sqrt{-G} G^{MN}$.
From (\ref{effectiveaction}) one sees that the fluctuations of the photon behave like a collection of
massless scalar fields propagating in a nonflat spacetime described by the metric $G_{MN}$, with a
position- and time-dependent gauge coupling given by  $g^2 = 1/V(f(t,r))$.

In the limit of condensation, $T^{MN}$ is given by (\ref{finalTmunu}), so that once the brane has formed
the action (\ref{effectiveaction}) reduces to a description of gauge fields propagating in a 
(3+1)-dimensional Minkowski space, with an additional component which couples the gauge fields to the 
tachyon matter density in the bulk.  In other words, the effective metric $G_{MN}$ starts off being
smooth throughout the bulk, but within the time $t_c$, its support collapses to become a delta function
$\delta^{(2)}(\vec x)$ in the relevant extra dimensions $\{r,\theta\}$.  

\section{Simplified Model of Reheating}

Notice that the extra bulk energy density $2 r \Sigma(r)$ in (\ref{finalTmunu}) yields a coupling between 
the gauge field $A^\subp_{\hat{\mu}}$ and the tachyon matter rolling in the bulk.
By changing coordinates which are 
comoving with the contraction of the vortex one can show that this coupling can be ignored in 
(\ref{effectiveaction}) since the gauge field is confined to the descendant brane.
The situation then closely resembles a gauge theory defined on a manifold in which a two-dimensional 
subspace which is shrinking with time. As a simplified model of the interaction we thus consider a massless
spin-1 field 
\begin{equation}
\label{toyaction}
S=-\frac{1}{4}\int  
    \sqrt{-g} g^{MA} g^{NB} F_{MN}^{\subp} F_{AB}^{\subp}
     \, d^{p+1} x
\end{equation}
propagating in a FRW-like background
\begin{equation}
\label{toymetric}
  g_{MN}dx^M dx^N = -dt^2 + R(t)^2\left(d\tilde r^2+\tilde r^2d\theta^2 \right) + 
                     \delta_{\hat{\mu}\hat{\nu}}dx^{\hat{\mu}} dx^{\hat{\nu}}.
\end{equation}
The coordinate $\tilde r$ in (\ref{toymetric}) is fixed with the expansion.

We will impose homogeneous boundary conditions at $\tilde{r}=1$ and take the scale factor in 
(\ref{toymetric}) to be
\[
  R(t) = \left \{ \begin{array}{ll} R_0 & \mbox{if $t < 0$}; \\
          R_0- t   & \mbox{if $0 \leq t \leq t_c$}; \\ 
          R_0- t_c = \epsilon  & \mbox{if $t>t_c$}  \end{array} \right.
\]
where $R_0$ represents the initial radial size of the extra dimensions and $\epsilon$ represents
the final state brane thickness. 
A shortcoming of this approximation is that $\partial_t R$ is discontinuous
at the interfaces, which leads to an ultraviolet divergence in the production of gauge bosons.  
The behavior of the actual $R(t)$ is smooth, and must yield a finite amount of particle production 
\cite{Reheating1} so we introduce a UV cut-off $p_{max} = \Lambda$ which we will assume is of order the 
inverse string length $l_s^{-1}$.  In addition, we will find a separate UV divergence in the particle
production in the limit as the vortex core thickness goes to zero ($\epsilon \rightarrow 0$).  This is 
presumably an artifact of the effective field theory which is not present in the full string theory, and we 
deal with it by  introducing the cutoff $\epsilon$ on the final radius of the defect core, which we will 
take to be of order $l_s$.

The equations of motion which follow from (\ref{toyaction}) can be solved exactly in each of the three
regions of the spacetime $t<0$, $0 \leq t \leq t_c$ and $t>t_c$.  In the asymptotic regions $t<0$ and
$t>t_c$ the gauge field is just the usual sum of plane waves with annihilation/creation operators
$\{a_{mn},a_{mn}^{\dagger}\}$ and $\{d_{mn},d_{mn}^{\dagger}\}$ respectively.  The indices $\{m,n\}$
label the KK modes of the photon.  The mass spectrum in the final state is of order 
$\epsilon^{-1} \sim l_s^{-1}$ which is the same order as the UV cut-off so we are obligated to neglect the 
massive modes for consistency.

\section{Energy Density of Produced Radiation}

We assume that the universe starts out in the vacuum which is annihilated by $a_{mn}$, $a_{mn} |0> =0$.
However, this vacuum is not annihilated by $d_{mn}$ which is a linear superposition of $a_{mn}$ and $
a_{mn}^{\dagger}$.  The coefficients in this superposition are the Bogoliubov coefficients which are
determined by matching the gauge field solutions in each of the three regions of the spacetime.  The spectrum
of particles observers in the future will see is then given by
\[
  N_{mn}(p) = <0|d_{mn}^{\dagger}(p) d_{mn}(p)|0>
\]
and the energy density of the produced radiation is 
\[
  \rho = 2\int \frac{d^3 p}{(2 \pi)^3} p N_{00} (p)
\]
where we have neglected all modes but the massless zero mode
and the multiplicative factor of $2$ corresponds to the two polarizations of the photon.  For large extra 
dimensions ($R_0 \gg l_s$) we find that the energy density takes the very simple form
\[
  \lim_{R_0 \to \infty} \rho = \frac{1}{8 \pi^2} \frac{\Lambda^2}{\epsilon^2}.
\]

\section{Efficiency of Reheating}

To quantify the efficiency of the reheating we need to determine how much energy is available to
produce the photons on the final-state 3-brane.  Initially the system consisted of a D5-brane plus
antibrane, whose 3D energy density was given by $2\tau_5 V_2$, where $\tau_5$ is the tension of a
D5-brane and $V_2$ is the volume of the compact 2-space $\{ r,\theta \}$ wrapped by the branes. The
final state consists of a D3-brane/antibrane with total tension $2\tau_3$.  Since reheating on each 
final-state brane should be equally efficient,  the 3D energy density
available for reheating on one them, which we call the critical energy density $\rho_c$,   is half
the difference between the initial and final tensions of the branes and antibranes: 
$\rho_c \equiv  \tau_5 V_2 - \tau_3$.
In the regime where $R_0 \gg l_s$ this simplifies to 
$l_s^4 \rho_c = \left(  16 \pi^3 \alpha(M_s) \right)^{-1}$
where $\alpha(M_s) \approx 1/25$ \cite{Jones} is the gauge coupling evaluated at the string scale.  We find 
that the criterion for efficient reheating becomes
\[
  \sqrt{\epsilon\over\Lambda} \lsim \left({2\pi\alpha(M_s)}\right)^{1/4} l_s  \cong 0.7\, l_s. 
\]
We find, then, that the reheating can be efficient taking $\Lambda^{-1}$, $\epsilon$ both close to
the string length.


\section{Conclusions and Discussion}

We have argued that our visible universe might be a codimension-two brane left
over from annihilation of a D5-brane/antibrane pair at the end of inflation.  In this picture, reheating
is due to production of standard model particles ({\it e.g.} photons) on the final branes, driven by
their couplings to the tachyon field which encodes the instability of the initial state as well
as the vortex which represents the final brane. We find that reheating can be
efficient, in the sense that a sizable fraction of the energy available from the unstable vacuum can
be converted into visible radiation, and not just gravitons.  

The efficiency of reheating is greatest  if the radius of compactification of the extra dimensions
is larger than the string length $l_s$.  The efficiency also depends on phenomenological
parameters we had to introduce by hand in  order to cut off ultraviolet divergences in the
calculated particle production rate: namely $\epsilon$, a nonvanishing radius for the final brane,
and $\Lambda$, an explicit cutoff on the momentum of the photons produced.  The latter must be
introduced to correct for discontinous time derivatives in our simplified model of the background
tachyon condensate; the actual behavior of the condensate corresponds to a cutoff of order
$\Lambda\sim 1/l_s$.  It is less obvious why the effective field theory treatment should give
divergent results as the thickness of the final brane ($2\epsilon$) goes to zero, but it seems clear
that a fully string-theoretic computation would give no such divergence, and therefore it is
reasonable to cut off the field-theory divergence at $\epsilon\sim l_s$.  Given only these mild
assumptions, our estimates predict that the fraction of the available energy which is converted
into visible radiation is $\rho/\rho_c \cong \pi\alpha(M_s) \cong 0.25$.  This simple estimate
counts only photons; in a more realistic calculation, it would be enhanced by the number of light
degrees of freedom which couple to the tachyon, which could be much greater 1.  Moreover it could
also be enhanced by the production of massive KK modes.


\section*{Acknowledgments}

We thank H. Stoica, R. Brandenberger and H. Tye for helpful comments and discussions.
This research was supported by the Natural Sciences and Engineering Research Council (NSERC) of Canada and 
Fonds qu\'eb\'ecois de la recherche sur la nature et les technologies (NATEQ).











\begin{thebibliography}{0}




\bibitem{Reheating2}
N.~Barnaby and J.~M.~Cline,
Phys.\ Rev.\ D {\bf 70}, 023506 (2004)
[arXiv:hep-th/0403223].

\bibitem{Reheating1}
J.~M.~Cline, H.~Firouzjahi and P.~Martineau,
JHEP {\bf 0211}, 041 (2002)
[arXiv:hep-th/0207156].

\bibitem{Sen}
A.~Sen,
Phys.\ Rev.\ D {\bf 68}, 066008 (2003)
[arXiv:hep-th/0303057].

\bibitem{Caustics}
N.~Barnaby,
JHEP {\bf 0407}, 025 (2004)
[arXiv:hep-th/0406120].
G.~N.~Felder and L.~Kofman,
Phys.\ Rev.\ D {\bf 70}, 046004 (2004)
[arXiv:hep-th/0403073].
G.~N.~Felder, L.~Kofman and A.~Starobinsky,
JHEP {\bf 0209}, 026 (2002)
[arXiv:hep-th/0208019].

\bibitem{Jones}
N.~Jones, H.~Stoica and S.~H.~H.~Tye,
JHEP {\bf 0207}, 051 (2002)
[arXiv:hep-th/0203163].




\end{thebibliography}
\end{document}